\documentclass[pra,aps]{revtex4}
\newcommand{\be}{\begin{equation}}
\newcommand{\ee}{\end{equation}}
\newcommand{\br}{\begin{eqnarray}}
\newcommand{\er}{\end{eqnarray}}
\newcommand{\nn}{\nonumber}
\newcommand{\bd}{\begin{displaymath}}
\newcommand{\ed}{\end{displaymath}}

\newcommand{\bib}{\bibitem}
\newcommand{\bfig}{\begin{figure}}
\newcommand{\efig}{\end{figure}}
\usepackage{ifvtex}
\usepackage{ifpdf}
\usepackage[dvips]{graphicx}
\usepackage{amsmath}
\usepackage{amssymb}
\usepackage{eucal}
\def\alf{\alpha}

\def\lam{\lambda}
\def\om{\omega}
\def\eps{\epsilon}
\def\rpar{\right)}
\def\lpar{\left(}
\def\rbk{\right]}
\def\lbk{\left[}

\def\lb{\label}
\def\im{{\rm i}}
\def\tr{\mbox{${\rm Tr}$}}

\def\ro{\mbox{\boldmath $\rho$}}
\def\sig{\mbox{\boldmath $\sigma$}}

\def\bsig{\mbox{\boldmath $\Sigma$}}

\def\bgam{\mbox{\boldmath $\gamma$}}
\def\sbgam{\mbox{\boldmath {\scriptsize $\gamma$}}}
\def\indn{\mbox{\tiny $N$}}
\def\cnot{\mbox{\tiny ${\rm CNOT}$}}
\def\tone{\mbox{\tiny ${\rm T}_{1}$}}
\def\ttwo{\mbox{\tiny ${\rm T}_{2}$}}
\def\hone{\mbox{\tiny ${\rm H}_{1}$}}
\def\htwo{\mbox{\tiny ${\rm H}_{2}$}}
\def\inde{\mbox{\tiny ${\rm E}$}}
\def\indi{\mbox{\tiny ${\rm I}$}}
\def\bu{\mbox{\boldmath ${\cal U}$}}
\def\bht{\mbox{\boldmath ${\cal H}$}}

\begin{document}
%
\title{Using continuous measurement to protect a universal set of quantum gates within a perturbed decoherence-free subspace}
\author{Paulo E. M. F. Mendon\c{c}a}
\affiliation{School of Physical Sciences, The University of Queensland, \\
             Queensland 4072, Australia \\
             E-mail address: mendonca@physics.uq.edu.au}
\author{Marcelo A. Marchiolli, Reginaldo d. J. Napolitano}
\affiliation{Instituto de F\'{\i}sica de S\~{a}o Carlos, Universidade de S\~{a}o Paulo, \\
             Caixa Postal 369, 13560-970, S\~{a}o Carlos, SP, Brazil \\
             E-mail address: marcelo$\_$march@bol.com.br and reginald@if.sc.usp.br}
\date{\today}
%
\begin{abstract}
\vspace*{0.1mm}
\begin{center}
\rule[0.1in]{142mm}{0.4mm}
\end{center}
We consider a universal set of quantum gates encoded within a perturbed decoherence-free subspace of four physical qubits. Using
second-order perturbation theory and a measuring device modeled by an infinite set of harmonic oscillators, simply coupled to the
system, we show that continuous observation of the coupling agent induces inhibition of the decoherence due to spurious perturbations.
We thus advance the idea of protecting or even creating a decoherence-free subspace for processing quantum information. \\
\vspace*{0.1mm}
\begin{center}
\rule[0.1in]{142mm}{0.4mm}
\end{center}
\end{abstract}
\maketitle
\section{Introduction}

The reality of entanglement and state superposition inherent to quantum mechanics has opened up astounding possibilities. In particular,
some problems whose known classical solutions require exponential-time algorithms can now, in principle, be solved in polynomial time,
if use is made of these quantum resources \cite{i1,i2,i3}. Unfortunately, the implementation of such an efficient algorithm encounters
an almost unsurmountable obstacle: the degrading and ubiquitous decoherence due to the unavoidable coupling with the environment
\cite{i4,i5}. However, if the agent coupling the quantum computer to its environment is degenerate, any quantum information processed
within the corresponding degenerate subspace does not decohere \cite{i6,i7,i8,i9,i10,i11,i12,i13,i14}. The recent experimental
verification and investigation of decoherence-free subspaces (DFS) \cite{i15,i16} have continuously increased their potential
application in quantum information processing \cite{i17,i18,i19,i20}. In practice, these DFS's are only approximate, because all
observables of a quantum system are ultimately coupling agents to the surroundings, some of these being incompatible with a particular
degenerate one that gives rise to a DFS. In this sense, any proposal intended to further protect such an approximate DFS is relevant to
the effort to make quantum computing a realistic endeavor.  

In any physical implementation of a quantum computer, the basic unit of information processed is the qubit, as the bit is the elementary
information processed by a classical computer. Whatever the representation of a qubit in a quantum computer, it should be operated
through unitary transformations, including the identity. Since the subspace spanned by a qubit is two dimensional, the manipulation of a
qubit must be represented by ${\rm SU(2)}$ transformations. The information processed by a quantum computer is, therefore, represented by
states of a tensor product of two-dimensional qubit subspaces. Universal quantum computation requires a set of quantum gates represented
by unitary transformations of two qubits at least \cite{i12,i21,i22}. 

An $N$-dimensional decoherence-free subspace, ${\cal H}_{\indn}$, is possible only when the agent coupling it to its surroundings is
an observable degenerate on ${\cal H}_{\indn}$ \cite{i9}. If one considers the implementation of a minimal set of universal quantum
gates operating on two qubits only, then there is no nontrivial degenerate observable available to establish a four-dimensional
decoherence-free subspace. Without loss of generality, in a system of four qubits (e.g., four $1/2$ spins), the nontrivial choice of
${\bf J}_{z}$ as the coupling observable allows us to construct two distinct four-dimensional decoherence-free subspaces associated with
the eigenvalues of ${\bf J}_{z}$ equal to $M_{z} = \pm \hbar$. Here we consider the four-dimensional subspace with $M_{z}=+\hbar$,
within which we assume that gate operations are described by the usual spin-boson Hamiltonian: 
\be
\lb{e1}
{\bf H}_{0}(t) = - \frac{\hbar}{2} \sum_{n=1}^{4} \lbk \mathbb{B}^{(n)}(t) \rbk^{\dagger} \bsig^{(n)} + \frac{\hbar^{2}}{4}
\sum_{n,m=1}^{4} \lbk \bsig^{(m)} \rbk^{\dagger} \mathbb{G}^{(mn)}(t) \, \bsig^{(n)}
\ee
where
\bd
\bsig^{(n)} = \lbk \begin{array}{c} \sig_{x}^{(n)} \\ \sig_{y}^{(n)} \\ \sig_{z}^{(n)} \end{array} \rbk \qquad 
\mathbb{B}^{(n)}(t) = \lbk \begin{array}{c} B_{x}^{(n)}(t) \\ B_{y}^{(n)}(t) \\ B_{z}^{(n)}(t) \end{array} \rbk \qquad \mbox{and} \qquad
\mathbb{G}^{(mn)}(t) = \lbk \begin{array}{ccc} G_{xx}^{(mn)}(t) & G_{xy}^{(mn)}(t) & G_{xz}^{(mn)}(t) \\
                                               G_{yx}^{(mn)}(t) & G_{yy}^{(mn)}(t) & G_{yz}^{(mn)}(t) \\
                                               G_{zx}^{(mn)}(t) & G_{zy}^{(mn)}(t) & G_{zz}^{(mn)}(t) \end{array} \rbk \; .
\ed
The operators $\sig_{j}^{(n)}$ with $j=x,y,z$ correspond to Pauli matrices for spin $n$, obeying the well-known ${\rm su}(2)$ commutation
relations. Furthermore, the real $(3 \times 1)$-matrix $\mathbb{B}^{(n)}(t)$ is associated with an external field acting locally on
qubit $n$, and the real $(3 \times 3)$-matrix $\mathbb{G}^{(mn)}(t)$ represents externally-controlled interactions between qubits $m$
and $n$.  

Decoherence is universal for real systems \cite{i23}. Essentially all the observables of a system are coupled to the environment,
rendering decoherence unavoidable. However, it has been proposed that continuous measurement of an observable protects (through the
quantum Zeno effect) states defined in the subspace associated to a degenerate eigenvalue of that observable \cite{i14}. In the present
work we show, using the simple system above as a paradigm, that a strong enough coupling of ${\bf J}_{z}$ to the environment effectively
creates ${\cal H}_{4}$. In other words, provided the initial states are chosen within $M_{z} = + \hbar$ and ${\bf H}_{0}(t)$ commutes
with ${\bf J}_{z}$, we show that the environment effects induced by couplings to system observables incompatible with ${\bf J}_{z}$ are
immaterial if the frequency of measurement of ${\bf J}_{z}$ is high enough.  

This work is organized as follows. Section II describes the construction of a universal set of quantum gates operating within a DFS. In
Section III we introduce a perturbation that degrades the original DFS and establish a scheme to inhibit the perturbing effects through
continuous measurement. Finally, the conclusion is presented in Section IV.  

\section{Quantum-gate construction in a decoherence-free subspace}

Universal quantum computation can be achieved by using any single member of the infinite class of universal sets of quantum gates. Any
gate capable of entangling two qubits, together with a minimal set of one-qubit gates form such a universal set \cite{i12,i21,i22}. For
our purposes, we choose the traditional set composed by the two-qubit controlled-not (CNOT) gate, the Hadamard gate, and the $\pi /8$
gate \cite{i24}.

To obtain the effective Hamiltonians for the mentioned universal set of quantum gates, we start from imposing that the general spin-boson
Hamiltonian ${\bf H}_{0}(t)$ commute with ${\bf J}_{z}$. After some algebra we obtain
\br
\lb{s1}
{\bf H}_{0}(t) &=& - \frac{\hbar}{2} \sum_{n=1}^{4} B_{z}^{(n)}(t) \, \sig_{z}^{(n)} + \frac{\hbar^{2}}{4} \sum_{n=1}^{3}
\sum_{m=n+1}^{4} \lbk G_{zz}^{(mn)}(t) \, \sig_{z}^{(m)} \sig_{z}^{(n)} + G_{xx}^{(mn)} (t) ( \sig_{x}^{(m)} \sig_{x}^{(n)} +
\sig_{y}^{(m)} \sig_{y}^{(n)} ) \right. \nn \\
& & + \left. G_{xy}^{(mn)}(t) ( \sig_{x}^{(m)} \sig_{y}^{(n)} - \sig_{y}^{(m)} \sig_{x}^{(n)} ) \rbk 
\er
where the independent coefficients $B_{z}^{(n)}(t)$, $G_{zz}^{(mn)}(t)$, $G_{xx}^{(mn)}(t)$, and $G_{xy}^{(mn)}(t)$ are arbitrary real
functions of time to ensure that ${\bf H}_{0}(t)$ be Hermitian. Below we show, by explicit construction, that the restriction of this
Hamiltonian to the four-dimensional subspace spanned by the eigenstates of ${\bf J}_{z}$ with eigenvalue $M_{z} = + \hbar$, 
${\cal H}_{4}$, is sufficiently general to form all the possible four-by-four Hermitian matrices. In this way, we equivalently prove
that universal quantum computation is possible within the decoherence-free subspace (DFS), ${\cal H}_{4}$, of our four-qubit model,
since the set of all Hermitian matrices produces the set of all unitary matrices through the exponential operation.

It is straightforward to show that the following choices of independent parameters form a complete set of Hermitian matrices within
${\cal H}_{4}$:
\be
\lb{s2}
{\bf H}_{\cnot} = \frac{\pi \hbar}{4 \tau} \, ( \sig_{z}^{(3)} + \sig_{z}^{(4)} - \sig_{x}^{(1)} \sig_{x}^{(2)} - \sig_{y}^{(1)}
\sig_{y}^{(2)} ) 
\ee
for the controlled-not gate,
\be
\lb{s3}
{\bf H}_{\tone} = - \frac{\pi \hbar}{8 \tau } \, ( \sig_{z}^{(3)} + \sig_{z}^{(4)} ) 
\ee
for the $\pi /8$ gate for the first qubit,
\be
\lb{s4}
{\bf H}_{\ttwo} = - \frac{\pi \hbar}{8 \tau} \, ( \sig_{z}^{(2)} + \sig_{z}^{(4)} ) 
\ee
for the $\pi /8$ gate for the second qubit,
\br
\lb{s5}
{\bf H}_{\hone} &=& \frac{\pi \hbar}{8 \tau} \left[ (2 - \sqrt{2}) ( \sig_{z}^{(1)} + \sig_{z}^{(2)} ) + (2 + \sqrt{2}) ( \sig_{z}^{(3)} 
+ \sig_{z}^{(4)} ) \right. \nn \\
& & - \left. \sqrt{2} \, ( \sig_{x}^{(1)} \sig_{x}^{(3)} + \sig_{y}^{(1)} \sig_{y}^{(3)} + \sig_{x}^{(2)} \sig_{x}^{(4)} + \sig_{y}^{(2)}
\sig_{y}^{(4)} ) \right]
\er
for the Hadamard gate for the first qubit, and
\br
\lb{s6}
{\bf H}_{\htwo} &=& \frac{\pi \hbar}{8 \tau} \left[ (2 - \sqrt{2}) ( \sig_{z}^{(1)} + \sig_{z}^{(3)} ) + (2 + \sqrt{2}) ( \sig_{z}^{(2)}
+ \sig_{z}^{(4)} ) \right. \nn \\
& & - \left. \sqrt{2} \, ( \sig_{x}^{(1)} \sig_{x}^{(2)} + \sig_{y}^{(1)} \sig_{y}^{(2)} + \sig_{x}^{(3)} \sig_{x}^{(4)} + \sig_{y}^{(3)}
\sig_{y}^{(4)} ) \right]
\er
for the Hadamard gate for the second qubit, where $\tau $ is a positive and real constant with dimension of time. We note that 
${\bf H}_{\ttwo}$ and ${\bf H}_{\htwo}$ are related with ${\bf H}_{\tone}$ and ${\bf H}_{\hone}$ through the interchange of the
superscripts $2$ and $3$. These Hamiltonians are just one possible set chosen, since the abundance of independent parameters renders the
associated linear system indeterminate. For this particular set of independent Hamiltonians to work according to conventional two-qubit
quantum computation, we define two abstract qubits spanning ${\cal H}_{4}$ according to the mapping:
\br
| 0,0 \rangle & \equiv & | \uparrow, \uparrow, \uparrow, \downarrow \rangle \nn \\
| 0,1 \rangle & \equiv & | \uparrow, \uparrow, \downarrow, \uparrow \rangle \nn \\
| 1,0 \rangle & \equiv & | \uparrow, \downarrow, \uparrow, \uparrow \rangle \nn \\
| 1,1 \rangle & \equiv & | \downarrow, \uparrow, \uparrow, \uparrow \rangle. \nn
\er
Now, by multiplying each of the above time-independent Hamiltonians by $- \im \tau / \hbar$ and exponentiating, we obtain, in matrix
format with respect to the ordered subspace basis $( | \uparrow, \uparrow, \uparrow, \downarrow \rangle, | \uparrow, \uparrow,
\downarrow, \uparrow \rangle, | \uparrow, \downarrow, \uparrow, \uparrow \rangle, | \downarrow, \uparrow, \uparrow, \uparrow \rangle )$:
\be
\lb{s7}
\begin{array}{ccc}
{\bf U}_{\cnot} = \exp (- \im \tau {\bf H}_{\cnot} / \hbar ) = \lbk \begin{array}{cccc}
1 & 0 & 0 & 0 \\
0 & 1 & 0 & 0 \\
0 & 0 & 0 & 1 \\
0 & 0 & 1 & 0 \end{array} \rbk 
\end{array}
\ee
for the controlled-not operation,
\be
\lb{s8}
\begin{array}{cc}
{\bf U}_{\tone} = \exp (- \im \tau {\bf H}_{\tone} / \hbar ) = \lbk \begin{array}{cccc}
1 & 0 & 0 & 0 \\
0 & 1 & 0 & 0 \\
0 & 0 & \exp \lpar \frac{\im \pi}{4} \rpar & 0 \\
0 & 0 & 0 & \exp \lpar \frac{\im \pi}{4} \rpar \end{array} \rbk
\end{array}
\ee
for the $\pi /8$ operation on the first abstract qubit,
\be
\lb{s9}
\begin{array}{cc}
{\bf U}_{\ttwo} = \exp (- \im \tau {\bf H}_{\ttwo} / \hbar ) = \lbk \begin{array}{cccc}
1 & 0 & 0 & 0 \\
0 & \exp \lpar \frac{\im \pi}{4} \rpar & 0 & 0 \\
0 & 0 & 1 & 0 \\
0 & 0 & 0 & \exp \lpar \frac{\im \pi}{4} \rpar \end{array} \rbk
\end{array}
\ee
for the $\pi /8$ operation on the second abstract qubit,
\be
\lb{s10}
\begin{array}{cc}
{\bf U}_{\hone} = \exp (- \im \tau {\bf H}_{\hone} / \hbar ) = \dfrac{1}{\sqrt{2}} \lbk \begin{array}{cccc}
1 & 0 & 1  & 0 \\
0 & 1 & 0  & 1 \\
1 & 0 & -1 & 0 \\
0 & 1 & 0  & -1 \end{array} \rbk 
\end{array}
\ee
for the Hadamard operation on the first abstract qubit, and
\be
\lb{s11}
\begin{array}{cc}
{\bf U}_{\htwo} = \exp (- \im \tau {\bf H}_{\htwo} / \hbar ) = \dfrac{1}{\sqrt{2}} \lbk \begin{array}{cccc}
1 & 1  & 0 & 0 \\
1 & -1 & 0 & 0 \\
0 & 0  & 1 & 1  \\
0 & 0  & 1 & -1 \end{array} \rbk 
\end{array}
\ee
for the Hadamard operation on the second abstract qubit. As mentioned above, these unitary transformations form a possible set of
operations suitable for universal computing.

Inspecting the right-hand sides of Eqs. (\ref{s2})-(\ref{s6}), we notice that the construction of the above model of universal
computation, within a decoherence-free subspace, does not require interactions represented by the tensor product of two spin-$1/2$
operators of different directions. We believe that being able to choose all the $G_{xy}^{(mn)}(t)$ equal to zero, and yet obtain
universal computing, might be relevant in the context of realistic implementation. In this section we have established, therefore, one
of the simplest protections against decoherence for a complete set of quantum-computation gates. In the next section we investigate how
robust this protection can be, by introducing an additional perturbing term to the Hamiltonian of a general quantum computation being
processed within the DFS.

\section{Measurement-induced inhibition of decoherence within a perturbed decoherence-free subspace}

A two-qubit quantum computation can be described as a sequence of quantum-gate operations on an input density matrix $\ro(0) = | \varphi
(0) \rangle \langle \varphi(0) |$, where $| \varphi(0) \rangle $ is the ket specifying the initial state of the two abstract qubits. If
the input state belongs to ${\cal H}_{4}$, then any computation involving the two abstract qubits can be performed by a sequence of the
universal operations described by (\ref{s2}) to (\ref{s6}), or, in a condensed form, by (\ref{s1}), where the time-dependent
coefficients vary as functions of time according to the specific computation, reproducing the required gate sequence.

We model the coupling with the environment by the product of ${\bf J}_{z}$ and observables of infinitely many harmonic oscillators representing the environment. By construction, the system Hamiltonian, at all times, satisfies
$[ {\bf H}_{0}(t),{\bf J}_{z} ] = 0$, since the computation is designed to occur entirely in ${\cal H}_{4}$. Now, it is obvious that any
perturbation added to ${\bf H}_{0}$, even if it acts only on the system, that couples ${\cal H}_{4}$ to its complement in the original
four-qubit Hilbert space, triggers the process of decoherence. Therefore, to simulate the possibility of this degradation of the
computation, it suffices to study the dynamics of the following Hamiltonian:
\be
\lb{t1}
{\bf H}(t) = {\bf H}_{0}(t) + {\bf H}_{\inde} + \lam {\bf J}_{z} \sum_{k} g_{k} ({\bf a}_{k} + {\bf a}_{k}^{\dagger}) + \eps 
{\bf J}_{x}
\ee
where $g_{k}$ is a coupling constant to the $k$th environmental degree of freedom, $\lam$ and $\eps$ are positive real constants
satisfying $\lam \gg \eps$, and the environment Hamiltonian is given by 
\be
\lb{t2}
{\bf H}_{\inde} = \sum_{k} \hbar \om_{k} \, {\bf a}_{k}^{\dagger} {\bf a}_{k} 
\ee
where ${\bf a}_{k}$ and ${\bf a}_{k}^{\dagger}$ are the annihilation and creation operators, respectively, of the $k$th quantum harmonic
oscillator, of frequency $\om_{k}$, representing one of the environmental degrees of freedom. We remark that here the meaning of the term ``environment" also possibly includes a measuring apparatus.  

Next we show that, as $\lam$ increases as compared to $\eps$, the degradation of the computation dynamics due to decoherence decreases
substantially, reaching a regime in which it can be safely neglected. Our interpretation of this fact parallels the ideas of \cite{i14}:
``the strong coupling to the environment functions as a continuous measurement of whether the perturbation takes the system state out of
${\cal H}_{4}$, protecting the dynamics against the perturbation." We are, therefore, benefiting from the quantum Zeno effect to induce
the inhibition of the decoherence process.

For convenience, let us define the unitary time-evolution operator
\be
\lb{t3}
\bu(t) \equiv \exp \lbk - \frac{\im t}{\hbar} \lpar {\bf H}_{\inde} + \lam {\bf J}_{z} \sum_{k} g_{k}
( {\bf a}_{k} + {\bf a}_{k}^{\dagger} ) \rpar \rbk \; .
\ee
Using the results of \cite{i25} and ${\bf J}_{\pm} = {\bf J}_{x} \pm \im {\bf J}_{y}$, we deduce that
\br
\lb{t4}
\bu^{\dagger}(t)\, {\bf J}_{x} \, \bu(t) &=& e^{\sbgam(t) {\bf J}_{z}} e^{- \im \alf(t) {\bf J}_{z}^{2}} \, {\bf J}_{x} \, e^{\im \alf(t)
{\bf J}_{z}^{2}} e^{- \sbgam(t) {\bf J}_{z}} \nn \\
&=& \frac{e^{\im \alf(t)}}{2} \lpar e^{[\sbgam(t) - 2 \im \alf(t) {\bf J}_{z} ]} \, {\bf J}_{+} + e^{- [ \sbgam(t)- 2 \im \alf(t) 
{\bf J}_{z} ]} \, {\bf J}_{-} \rpar
\er
where
\bd
\alf(t) = \sum_{k} \lpar \frac{\lam g_{k}}{\om_{k}} \rpar^{2} \lbk (\om_{k} t) - \sin(\om_{k} t) \rbk \qquad \qquad
\bgam(t) = \sum_{k} [ f_{k}(t) {\bf a}_{k}^{\dagger} - f_{k}^{\ast}(t) {\bf a}_{k} ]
\ed
with
\bd
f_{k}(t) = - \frac{\lam g_{k}}{\om_{k}} [ 1 - e^{\im (\om_{k} t)} ] \; .
\ed
Furthermore, let us denote by ${\bf U}_{0}(t)$ the unitary operator that satisfies the evolution equation
\bd
\im \hbar \, \frac{d{\bf U}_{0}(t)}{dt} = {\bf H}_{0}(t) {\bf U}_{0}(t)
\ed
with ${\bf U}_{0}(0) = {\bf 1}$, where ${\bf 1}$ is the identity operator on ${\cal H}_{4}$. Because ${\bf H}_{0}(t)$ commutes with
${\bf J}_{z}$, it follows that ${\bf U}_{0}(t)$ commutes with $\bu(t)$, and the Hamiltonian for the qubits and its environment in the
interaction picture is written as
\be
\lb{t5}
\bht(t) = \eps \, {\bf U}_{0}^{\dagger}(t) \lbk \bu^{\dagger}(t) \, {\bf J}_{x} \, \bu(t) \rbk {\bf U}_{0}(t) \; .
\ee
%

\begin{figure}[!t]
\centering
\begin{minipage}[b]{0.60\linewidth}
\includegraphics[width=\linewidth]{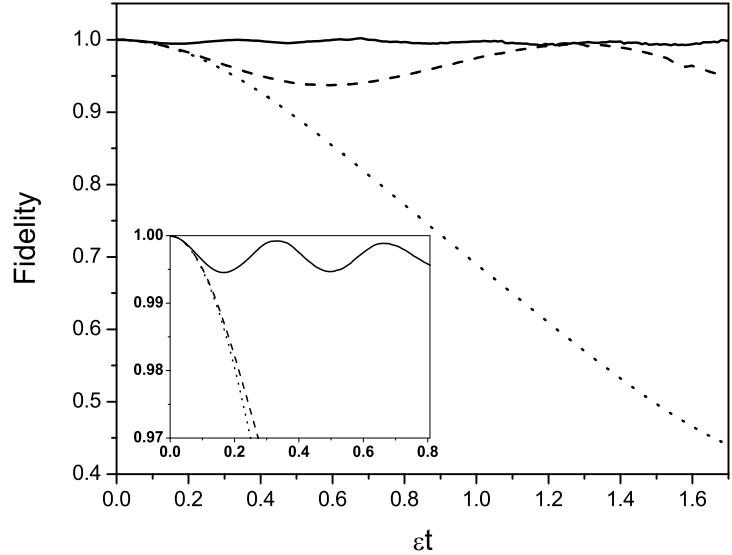}
\end{minipage}
\caption{Fidelity as a function of dimensionless time $\eps t$ for different values of $\Lambda$. For this particular illustration, we
take $\nu_{c} = 10^{5}$ and consider the following values of $\Lambda$: 2000 (solid line), 1150 (dashed line), and 800 (dotted line).
In this figure, the gate-operation time is $\tau$, $\eps$ being such that $\eps \tau = 1$. The inset shows the details of the fidelity
as a function of $\eps t$, where we can see that this function never exceeds one preserving the established superior limit, as it 
should.}
\end{figure}
The fidelity of the computing process is given by \cite{i24}
\be
\lb{t6}
{\cal F}(t) = \tr_{\inde} \lbk \langle \varphi (0) | \ro_{\indi}(t) | \varphi (0) \rangle \rbk
\ee
where the trace is taken over the environmental degrees of freedom and $\ro_{\indi}(t)$ is the density matrix of the qubits and its
environment in the interaction picture, i.e., $\ro _{\indi}(t) = | \Psi_{\indi}(t) \rangle \langle \Psi_{\indi}(t) |$, with 
$| \Psi_{\indi}(t) \rangle$ satisfying the Schr\"{o}dinger equation
\be
\lb{t7}
\im \hbar \, \frac{d | \Psi_{\indi}(t) \rangle}{dt} = \bht(t) | \Psi_{\indi}(t) \rangle \; .
\ee
Equation (\ref{t7}) can be solved iteratively in the usual approach of perturbation theory and it is easy to show that the first-order
term does not contribute to the fidelity. To obtain the second-order contribution, we need to calculate the quantity $\langle \varphi (0)
| \bht(t_{1}) \bht(t_{2}) | \Psi_{\indi}(0) \rangle$. Of course, the qubits are initially prepared in a pure state and $| \Psi_{\indi}
(0) \rangle$ is initially factored so that $| \Psi_{\indi}(0) \rangle = | \varphi (0) \rangle \otimes | {\rm E} \rangle$, where
$| {\rm E} \rangle $ is the initial state of the environment. Hence, to evaluate the second-order perturbation contribution to the
fidelity we need to calculate the quantity $\langle \Phi (t_{1}) | \Phi (t_{2}) \rangle$, where
\bd
| \Phi(t) \rangle = \bht(t) | \varphi (0) \rangle \; .
\ed

Now, let us define the auxiliary quantities $| \chi _{\pm}(t) \rangle \equiv {\bf U}_{0}^{\dagger}(t) \, {\bf J}_{\pm} \, {\bf U}_{0}(t)
| \varphi (0) \rangle$. Because $| \varphi (0) \rangle$ belongs to ${\cal H}_{4}$, it follows that ${\bf J}_{z} |\chi _{-}(t)\rangle = 0$
and ${\bf J}_{z} | \chi_{+} (t) \rangle = 2 | \chi_{+}(t) \rangle$. Therefore, it is easy to check that
\br
\lb{t8}
\langle \Phi (t_{1}) | \Phi (t_{2}) \rangle &=& \Gamma_{++}(t_{1},t_{2}) \, e^{-\sbgam(t_{1})} e^{\sbgam(t_{2})} +
\Gamma_{+-}(t_{1},t_{2}) \, e^{-\sbgam(t_{1})} e^{-\sbgam(t_{2})} \nn \\
& & + \, \Gamma_{-+}(t_{1},t_{2}) \, e^{\sbgam(t_{1})} e^{\sbgam(t_{2})} + \Gamma_{--}(t_{1},t_{2}) \, e^{\sbgam(t_{1})}
e^{-\sbgam(t_{2})}
\er
where we define the c-number functions as
\br
\Gamma_{++}(t_{1},t_{2}) &=& \frac{1}{4} \, e^{3 \im [ \alf(t_{1}) - \alf(t_{2}) ]} \, \langle \chi_{+}(t_{1}) | \chi_{+}(t_{2}) \rangle 
\nn \\
\Gamma_{+-}(t_{1},t_{2}) &=& \frac{1}{4} \, e^{\im [ 3 \alf(t_{1}) + \alf(t_{2}) ]} \, \langle \chi_{+}(t_{1}) | \chi_{-}(t_{2}) \rangle 
\nn \\
\Gamma_{-+}(t_{1},t_{2}) &=& \frac{1}{4} \, e^{- \im [ \alf(t_{1}) + 3 \alf(t_{2}) ]} \, \langle \chi_{-}(t_{1}) | \chi_{+}(t_{2})
\rangle \nn \\
\Gamma_{--}(t_{1},t_{2}) &=& \frac{1}{4} \, e^{- \im [ \alf(t_{1}) - \alf(t_{2}) ]} \, \langle \chi_{-}(t_{1}) | \chi_{-}(t_{2}) \rangle
\nn \; .
\er
To calculate the contribution of order $\eps^{2}$ to the fidelity, the trace over the environmental degrees of freedom requires the
evaluation of expectation values like $\langle {\rm E} | e^{- \sbgam(t_{1}) } e^{\sbgam(t_{2}) } | {\rm E} \rangle$. For this purpose,
we first observe that $\bgam(t)$ has a mathematical structure analogous to the argument of the displacement operator ${\bf D}(z_{k}) =
\exp ( z_{k} {\bf a}_{k}^{\dagger} - z_{k}^{\ast} {\bf a}_{k} )$ for a particular $k$-oscillator belonging to the environment. The 
second step consists in the expansion of $| {\rm E} \rangle$ in a convenient basis which permits us to include a wide class of 
environmental states (e.g., a Fock basis expansion with arbitrary coefficients). Thus, the expectation values can be promptly calculated
and their final result are proportional to $\exp(- \lam^{2} {\rm F})$, where F is a real function of $t_{1}$ and $t_{2}$ \cite{i26}. It
is important to mention that contributions of higher order than $\eps^{2}$ present the same factor in the calculations of 
expectation values and consequently similar analysis can be applied, implying that the fidelity tends toward unity as $\lam$ increases.

To ilustrate these calculations, for computational convenience, we have assumed a continuum-mode approximation with a non-ohmic spectral
density $R(\om) = (\om^{2} / 2 \om_{c}^{3}) e^{- \om / \om_{c}}$, cutting off exponentially as $\om$ gets greater than the cut-off
frequency $\om_{c}$. Furthermore, for the sake of simplicity, we also take a constant coupling $g_{k} = 1$ for all $k$, and a nontrivial
initial superposition state given by $| \varphi (0) \rangle = (1/ \sqrt{2}) (|1,0 \rangle - | 0,0 \rangle)$ in the abstract basis. The
fidelity (\ref{t6}) for this case is shown in figure 1, where we define the dimensionless parameters $\Lambda = \lam / \eps$ and
$\nu_{c} = \om_{c} / \eps$. We confirm that, in this simple ilustrative example, as $\Lambda$ increases the fidelity function oscillates
tending to unity and this fact corroborates our general model. 

\section{Conclusion}

Nowadays, the implementation of quantum gates within decoherence-free subspaces is one of the fundamental strategies in the development
of a realistic quantum-computer technology. The current experimental investigations have only considered existing DFS's, without
attempting to protect the quantum-information processing \cite{i15,i16,i17,i18,i19,i20}. Intending to improve this scenario, our present
proposal advances the idea of inhibiting the degradation of the gate operation within the DFS by continuous measurement. Although we
have illustrated this idea through a simplistic model, within a perturbation-theory context, we strongly believe that our results 
are not particular. A non-perturbative generalization of the present approach, independent of the model used to describe the environment
and its coupling to the system, is currently under our scrutiny.

\section*{Acknowledgments}

This work has been supported by Funda\c{c}\~{a}o de Amparo \`{a} Pesquisa do Estado de S\~{a}o Paulo (FAPESP), Brazil, projects $\sharp$
01/11562-2 (PEMFM), $\sharp$ 01/11209-0 (MAM) and $\sharp$ 00/15084-5 (RJN). We also aknowledge supports from the Millennium Institute
for Quantum Information - Conselho Nacional de Desenvolvimento Cient\'{\i}fico e Tecnol\'{o}gico (CNPq), Brazil.


\end{document}